\def\vecq{\ifmmode \vec q \else $\vec q\,$\fi}
\def\dhe{\ifmmode {\rm (d,^3He)} \else (d,$^3$He)\fi}
\def\eep{\ifmmode {\rm (e,e'p)} \else (e,e$'$p)\fi}
\def\kseven{\ifmmode ^{47}\rm K \else $^{47}$K\fi}
\def\caeight{\ifmmode ^{48}\rm Ca \else $^{48}$Ca\fi}
\def\halfpl{\ifmmode 1/2^+ \else 1/2$^+$\fi}
\def\threepl{\ifmmode 3/2^+ \else 3/2$^+$\fi}
\def\fivepl{\ifmmode 5/2^+ \else 5/2$^+$\fi}
\def\sevenmin{\ifmmode 7/2^- \else 7/2$^-$\fi}
\def\twos{\ifmmode {\rm 2s}_{1/2} \else 2s$_{1/2}$\fi}
\def\dthree{\ifmmode {\rm 1d}_{3/2} \else 1d$_{3/2}$\fi}
\def\newem{$E_{m}$}
\def\newpm{$\vec {p}_m$}
\def\newfseven{1f$_{7/2}$}
\def\newdfive{1d$_{5/2}$}

\documentstyle[epsf]{elsart}

\begin{document}

\begin{frontmatter}

\title{A consistent analysis of \eep\ and \dhe\ experiments} 

\author[NIKHEF]{G.J.~Kramer\thanksref{JAPAN}}
,
\author[VUA,NIKHEF]{H.P.~Blok} 
and 
\author[NIKHEF]{L.~Lapik\'as} 

\address[NIKHEF]{National Institute for Nuclear Physics and 
High-Energy Physics (NIKHEF),\\
P.O. Box 41882, 1009 DB Amsterdam, The Netherlands}
\address[VUA]{Department of Physics and Astronomy, Vrije 
Universiteit,\\
de Boelelaan 1081, 1081 HV Amsterdam, The Netherlands.}
\thanks[JAPAN]{present address: Ê
Naka Fusion Research Establishment,
Japan Atomic Energy Research Institute,
Naka-machi, Naka-gun,
Ibaraki-ken,
Japan}

%{\bf version 2.5, 000721/LL}

\begin{abstract}
The apparent discrepancy between spectroscopic factors obtained in 
\eep\ and \dhe\ experiments is investigated.  This is performed first 
for \caeight\eep\ and \caeight\dhe\ experiments and then for other 
nuclei.  It is shown that the discrepancy disappears if the \dhe\ 
experiments are re-analyzed with a non-local finite range DWBA 
analysis with a bound-state wave function that is obtained from \eep\ 
experiments.
\end{abstract}


\end{frontmatter}

%%%%%%%%%%%%%%%%%%%%%%%%%%%%%%%%%%%%%%%%%%%%%%%%%%%%%%%%%%%%%%%%%

\section{Introduction}
\label{sec:intro}

\begin{figure}
\centerline{\epsfysize=7.5cm \epsfbox{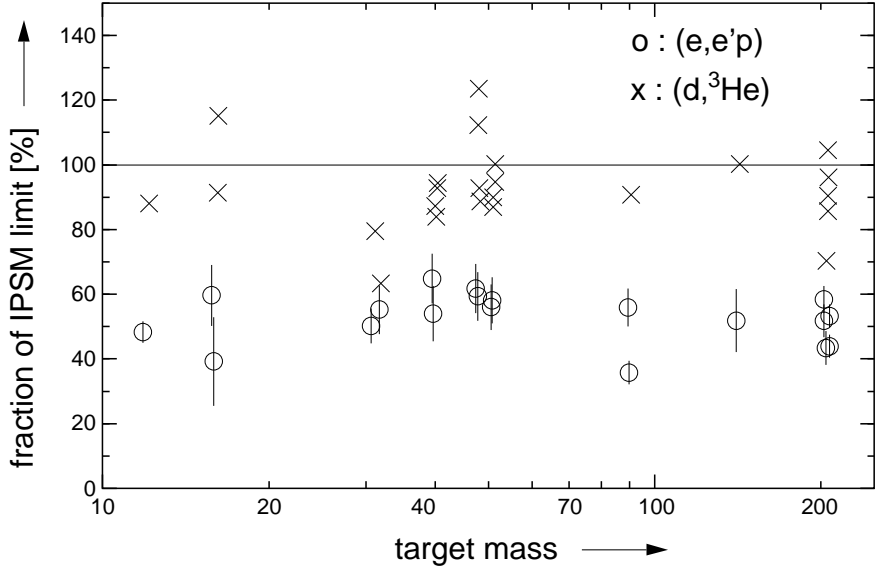} }
	\caption{Spectroscopic strength relative to the 
	Independent-Particle Shell Model limit for valence orbitals as a
	function of the mass number according to literature values for 
	\dhe\ and \eep\ experiments.  For references see  
	Table~\ref{tab:allspecfac}.  }
\label{fig:saval}
\end{figure}

Spectroscopic factors deduced from \eep\ reactions (see 
Fig.~\ref{fig:saval}) are found to be substantially lower 
\cite{blok87,her88,Lap93} than the sum-rule limit given by the 
independent-particle shell model (IPSM).  In contrast to this, 
experiments with hadronic probes such as the \dhe\ reaction, generally 
find spectroscopic factors that are close to values predicted by the 
IPSM. However, there is a strong model dependence in the extraction of 
these spectroscopic factors from transfer reactions (see e.g.  Ref.  
\cite{kra88a} and references therein).  In this paper it will be 
investigated to what extent this model dependence can account for the 
apparent discrepancy between reported spectroscopic factors derived 
from \eep\ and from \dhe\ experiments.

Modern nuclear-structure calculations 
\cite{pan84,hsi84,mahs88,piep90,Pud97,Geu96,Rij92,Rij96} predict 
occupations for valence orbitals in the range of 60 to 90~\%
of the IPSM limit.  The precise value and the spreading of the 
strength depend sensitively on the amount of short- and long-range 
correlations included in the calculation.  Recently, it was 
demonstrated \cite{lap99} for the nucleus $^{7}$Li that structure 
calculations based on a realistic nucleon-nucleon potential are indeed 
able to describe accurately the momentum distributions and 
spectroscopic factors measured with the reaction \eep.  To put such 
calculations to a further test for other nuclei it is necessary to 
avail of accurate absolute spectroscopic factors.  In this respect the 
existing discrepancy between spectroscopic factors deduced from the 
\eep\ and \dhe\ reactions needs a detailed investigation.  It is the 
aim of the present paper to carry out such a study and to provide a 
consistent set of spectroscopic factors extracted from both the \eep\ 
and the \dhe\ reaction.

In Section~\ref{sec:reactions} the Coulomb Distorted Wave Impulse 
Approximation (CDWIA), which is used in the analysis of the \eep\ 
experiments, and the Distorted Wave Born Approximation (DWBA) method, 
used for the \dhe\ reaction, are reviewed with special emphasis on the 
sensitivities of the spectroscopic factors to the various 
approximations made.  In Section~\ref{sec:bswf} it is investigated, 
which part of the bound-state wave function (BSWF) is probed by the 
\eep\ and \dhe\ reactions, in order to understand the model 
sensitivity arising from the shape of the BSWF. In Section~ 
\ref{sec:cadata} one \caeight\eep\ \cite{kra88b} and two \caeight\dhe\ 
\cite{banks85,mats84} data sets are used for a detailed comparison 
between the \eep\ and \dhe\ spectroscopic factors.  In Section~ 
\ref{sec:reanalysis} a re-analysis of \dhe\ data sets for other nuclei 
is made, in which non-locality and finite range corrections are 
included and BSWF's deduced from \eep\ experiments are used.  
Conclusions are drawn in Section~\ref{sec:conclusion}.

%%%%%%%%%%%%%%%%%%%%%%%%%%%%%%%%%%%%%%%%%%%%%%%%%%%%%%%%%%%%%%%%%

\section{Description of the reactions \eep\ and \dhe}
\label{sec:reactions}

\subsection{The \eep\ reaction}

In \eep\ experiments the energy ${e_i}$ and momentum $\vec k_i$ of the 
initial electron, and the energies and momenta of the final electron 
and knocked-out proton, denoted by ${e}_f$, $\vec k_f$ and ${E}_p$, 
$\vec k_p$, respectively, are measured.  The energy and momentum 
transferred by the scattered electron are denoted by~: $\omega = {e}_i 
- {e}_f$ and $\vec q = \vec k_i - \vec k_f$.  From energy and momentum 
conservation the missing energy \newem\ and missing momentum \newpm\ 
are determined~:
\begin{eqnarray}
&&{E}_m = {e}_i - {e}_f - {E}_p - {T}_{A-1} = 
\omega - {E}_p - {T}_{A-1} \nonumber \\
&&\vec {p}_m = \vec k_p^{lab} - (\vec k_i - \vec k_f) = 
\vec k_p - \vec q,
\label{eq:empmdef}
\end{eqnarray}
where $T_{A-1}$ is the kinetic energy of the residual nucleus.

The missing energy is the energy required to separate the struck 
proton from the target nucleus, where the final nucleus is left in the 
ground-state or in one of its excited states.  The missing momentum 
is, according to the definition of Eq.~(\ref{eq:empmdef}), the proton 
momentum in the nucleus just before the reaction provided that there 
is no further interaction between the incoming electron and the 
initial nucleus and the outgoing electron and proton and the final 
nucleus.

In the plane-wave impulse approximation (PWIA, see below) the \eep\ 
cross section can be written as~:
\begin{equation}
{{\rm d}^6\sigma \over {\rm d} E_{e\prime} {\rm d} \Omega_{e\prime} 
{\rm d} E_{p} {\rm d} \Omega_{p} } =
{k}\sigma_{ep}S(E_{m},\vec {p}_m),
\label{eq:sigdef}
\end{equation}
where the left-hand side represents the measured \eep\ cross section, 
$k\sigma_{ep}$ a kinematic factor times the elementary electron-proton 
cross section and $S$(\newem,\newpm) the spectral function 
\cite{frul84,kel96,diep90,bof93}.  The spectral function is the joint 
probability of finding a proton with momentum \newpm\ and binding 
energy \newem\ inside the nucleus.  For a transition leading to a 
discrete state at \newem=$E_{tr}$, the spectral function is written as 
the momentum distribution $\rho$(\newpm) times a delta function for 
the energy~:
\begin{equation}
S(E_{m},\vec {p}_m) = \rho(\vec {p}_m) \delta(E_{m}-{E}_{tr}).
\label{eq:spfdef}
\end{equation}

The spectral function as given in Eq.~(\ref{eq:sigdef}) cannot be 
determined experimentally because the outgoing proton interacts 
strongly with the final nucleus (this is often called the Final State 
Interaction, FSI).  Moreover the factorization of the six-fold 
differential cross section into an elementary electron-proton cross 
section times the spectral function does not hold any more due to the 
FSI and Coulomb distortions of the electron waves.

In the following a theoretical basis similar to the one used in the 
description of the \dhe\ reaction is given to compute the \eep\ cross 
section, in which the interaction between the participating particles 
is taken into account.  The influence of various approximations made 
in this analysis is also investigated in order to reveal the origin of 
the model uncertainties on the extracted observables.  Since we want 
to keep the formulae transparent the angular momentum and spin parts 
are not given in this article.

The basis of the theoretical description of the reaction~: $A+e_i 
\rightarrow B+p+e_f $ is the T-matrix formalism 
\cite{satch80,aust70,jack70}.  This T-matrix in the prior form is 
defined in the following way~:
\begin{equation}
T_{if} ~=~ < \Psi^{(-)}_f ~| V_i - U_i |~\psi_A~\varphi^{(+)}_i >,
\label{eq:tmatdef}
\end{equation}
where $\varphi^{(+)}_i(\vec k_i,\vec r\,)$ is the incoming distorted 
electron wave with outgoing-wave boundary conditions, $\psi_A$ the 
wave function of the target nucleus, $V_i$ the total interaction 
between the incoming electron and the nucleus, from which $U_i$, the 
potential used to generate the distorted wave $\varphi_i$, is 
excluded, and $\Psi^{(-)}_f$ is the exact final-state wave function of 
the electron-proton-residual-nucleus system obeying incoming-wave 
boundary conditions.  The distorting potential $U_i$ is usually taken 
to be the Coulomb potential arising from a uniformly charged sphere.

\begin{figure}
\centerline{\epsfysize=3cm \epsfbox{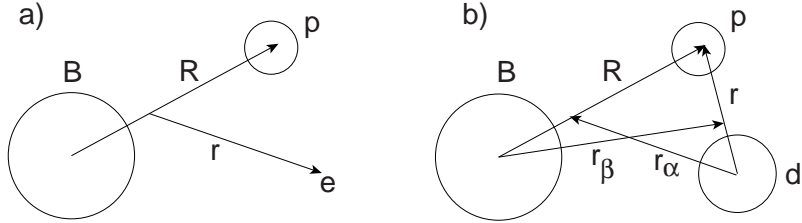} }
\caption{ The geometry used in the description of the \eep\ reaction 
(a) and the \dhe\ reaction (b).  }
\label{fig:reaction}
\end{figure}

In the distorted-wave approximation the exact final-state wave 
function is approximated by the product of an internal wave function 
for the residual nucleus $\psi_B$, a distorted outgoing electron-wave 
$\varphi^{(-)}_f(\vec k_f,\vec r\,)$, and the distorted outgoing 
proton-wave $\chi^{(-)}_p(\vec k_p,\mu \vec R\,)$ (with 
$\mu=(A-1)/A$).  Here the displacement of the knocked-out proton from 
the residual nucleus $B$ is denoted by $\vec R$, while $\vec r\,$ is 
the displacement of the electron from the center of mass of the 
residual-nucleus plus proton final system (see 
Fig.~\ref{fig:reaction}a).  Under these assumptions the distorted wave 
transition amplitude becomes~:
\begin{equation}
T^{DW}_{if} =~
< \varphi^{(-)}_f(\vec k_f,\vec r\,)~
\chi^{(-)}_p(\vec k_p,\mu \vec R\,)~
\psi_B~| V_i - U_i |~\psi_A~
\varphi^{(+)}_i(\vec k_i,\vec r\,) >.
\label{eq:tmat}
\end{equation}
For light and medium-heavy nuclei the effects of the electron 
distortions are small \cite{rosen80}; for heavy nuclei, however, these 
effects become sizable.  Because of the long range of the Coulomb 
potential it is difficult to include electron distortions in 
theoretical codes that compute \eep\ momentum distributions.  Up to 
now two approaches have been followed to deal with these distortions.  
In the simplest one, the Effective Momentum Approximation 
(EMA)~\cite{YenB65}, the electron momenta are replaced by effective 
ones.  A more precise treatment of the electron distortions is 
employed in the work of Giusti and Pacati \cite{gius87,gius88}, in 
which the eikonal approach is used to expand the electron waves in 
powers of $Z \alpha$ ($Z$ the nuclear charge and $\alpha$ the 
fine-structure constant).  For medium heavy nuclei such as the calcium 
isotopes this approximation is accurate enough as all higher order 
terms have negligible influence on the calculated cross sections 
\cite{gius88}.  Recently, full relativistic calculations 
\cite{jin93,udias93,jin92} have been used to analyze momentum 
distributions measured with the reaction \eep.  The deduced 
spectroscopic factors are different by up to 10\% from those resulting 
from a non-relativistic analysis.  However, in these analyses 
relativistic Hartree-Fock wave functions are used for the BSWF, which 
not always provide a satisfactory description of the experimental 
momentum distributions.  Moreover, the use of relativistic optical 
potentials at the low proton energies ($T_{p} \le$ 100 MeV) employed 
in the presently discussed experiments may be questionable.  We 
therefore limit the present analysis to the non-relativistic approach.

The distorted proton wave, $\chi^{(-)}_p(\vec k_p,\mu \vec R\,)$, is 
usually chosen to be the solution of the Schr\"odinger equation with 
the optical potential that describes elastic proton scattering off the 
final nucleus, $B$.  The parameterization of the optical potential is 
not unique; using a different parameterization to generate the 
distorted proton waves results in waves identical at large distances 
but different in the nuclear region where the \eep\ reaction takes 
place.  The effect of different parameterizations on the value of the 
extracted spectroscopic factors has been investigated for the reaction 
$^{51}$V\eep\ \cite{her88}.  A model uncertainty of about 6~\% was
found there due to the treatment of the final-state interaction.

The optical potential used for the generation of the distorted waves 
as well as the binding potential for the proton are local potentials.  
However, for fundamental reasons this potential is expected to be 
non-local.  Perey \cite{perey63} has pointed out that the wave 
function of a non-local potential is systematically smaller in the 
nuclear interior than the wave function of the local potential that 
gives an equivalent description of the elastic scattering process.  
This non-locality correction can be taken into account effectively by 
multiplying the local wave function with the factor~:
\begin{equation}
F(r) = \bigl(1- { {\mu_p \beta^2} \over {2 \hbar^2} } 
U_L(r)\bigr)^{-1/2},
\label{eq:perey}
\end{equation}
where $\mu_p={{A-1} \over A}m_p$ is the reduced proton mass, $U_L(r)$ 
the local optical potential and $\beta$ the range of the non-locality.  
The non-locality correction affects the distorted waves in the region 
where the potential is significantly different from zero.  This is 
also the region where the reaction takes place so that the 
non-locality correction also affects the spectroscopic factors 
determined from knock-out reactions.

The T-matrix element $T^{DW}_{if}$ can be written in a way that 
explicitly shows the nuclear-structure part~:
\begin{eqnarray}
T^{DW}_{if} =~\int d\vec r \int d\vec R &
\varphi^{(-)*}_f(\vec k_f,\vec r\,)~\chi^{(-)*}_p(\vec k_p,
\mu \vec R\,) \times \nonumber \\
&< \psi_B | V_i - U_i | \psi_A >\varphi^{(+)}_i(\vec k_i,\vec r\,) .
\label{eq:tmat2}
\end{eqnarray}
The matrix element {$< \psi_B | V_i - U_i | \psi_A >$} contains the 
nuclear-structure information.  It involves integration over all 
internal coordinates $\xi_B$, independent of $\vec r$ and $\vec R$.  
The potential $V_i$ describes the total interaction between the 
electron and the nucleus, whereas in the distorting potential $U_i$ 
the part of the interaction leading to the \eep\ channel is excluded.  
In this way the difference $V_i - U_i$ is the interaction between the 
electron and the struck proton~:
\begin{equation}
V_i - U_i = V_{ep}( \vec r,\mu \vec R\,).
\label{eq:vminu}
\end{equation}
Since $V_{ep}$ is independent of the internal coordinates of $\psi_B$, 
the nuclear matrix element can be factorized into $V_{ep}$ and the 
nuclear overlap integral~:
\begin{equation}
< \psi_B | V_i - U_i | \psi_A >~=~
V_{ep}(\vec r,\mu \vec R\,) 
\int d \xi_B~
\psi^*_B (\xi_B)
\psi_A (\xi_B,\vec R\,).
\label{eq:overlap}
\end{equation}
The integral is usually expanded into single-particle states 
\cite{satch80} :
\begin{equation}
\int d \xi_B
\psi^*_B (\xi_B)
\psi_A (\xi_B,\vec R\,) =
\sum_{n \ell j m} \!<J_B j M_B m | J_A M_A>\!
\sqrt {S_{n \ell j}}~\phi_{n \ell j m} (\vec R\,),
\label{eq:overlapexp}
\end{equation}
where $<\!J_B j M_B m | J_A M_A\!>$ is a Clebsch-Gordan coefficient, 
$\sqrt {S_{n \ell j}}$ the spectroscopic amplitude and $\phi_{n \ell j 
m} (\vec R\,)$ a normalized single particle wave function, usually 
referred to as the bound state wave function (BSWF).  Substituting the 
foregoing two expressions into the transition amplitude yields~:
\begin{eqnarray}
T^{CDWIA}_{if} = &
\sum_{n \ell j m} \!<J_B j M_B m | J_A M_A>\! \sqrt {S_{n \ell j}}
\int d\vec r \int d\vec R
\varphi^{(-)*}_f(\vec k_f,\vec r\,) \times \nonumber \\
& \chi^{(-)*}_p(\vec k_p,\mu \vec R\,) 
V_{ep}(\vec r,\mu \vec R\,)~
\phi_{n \ell j m} (\vec R\,)~
\varphi^{(+)}_i(\vec k_i,\vec r\,), 
\label{eq:tcdwia}
\end{eqnarray}
which is the Coulomb Distorted Wave Impulse Approximation (CDWIA) 
amplitude.  In the code DWEEPY~\cite{gius88} that was used to 
calculate the momentum distributions presented in 
Section~\ref{sec:cadata}, this expression is evaluated together with 
the angular momentum and spin parts, which are not shown in 
Eq.~(\ref{eq:tcdwia}).  From Eq.~(\ref{eq:tcdwia}) the Distorted-Wave 
Impulse Approximation (DWIA) amplitude is obtained by replacing the 
electron waves by plane waves~:
\begin{eqnarray}
T^{DWIA}_{if} = &
\sum_{n \ell j m} \!<J_B j M_B m | J_A M_A>\! \sqrt {S_{n \ell j}}
\int d\vec r \int d\vec R
\exp( -i \vec k_f \!\cdot\! \vec r\,) \times \nonumber \\
& \chi^{(-)*}_p(\vec k_p,\mu \vec R\,) 
V_{ep}(\vec r,\mu \vec R\,)~
\phi_{n \ell j m} (\vec R\,)~
\exp( i \vec k_i \!\cdot\! \vec r\,).
\label{eq:tdwia}
\end{eqnarray}
In order to gain some further insight into Eq.~(\ref{eq:tdwia}) and to 
make the connection with the PWIA expression the Coulomb potential is 
now used for the interaction~: $V_{ep} = \alpha / |\mu \vec R - \vec 
r\,|$.  With this potential the integration over $\vec r\,$ can be 
performed~:
\begin{eqnarray}
T^{DWIA}_{if} = &
\sum_{n \ell j m} \!<J_B j M_B m | J_A M_A>\! 
{{4\pi \alpha} \over {\vecq\,^2}} \sqrt {S_{n \ell j}} 
\times \nonumber \\
& \int d\vec R~
\exp( -i \vecq \!\cdot\! \mu \vec R\,)~
\chi^{(-)*}_p(\vec k_p,\mu \vec R\,)~
\phi_{n \ell j m} (\vec R\,),
\label{eq:tdwiaprop}
\end{eqnarray}
where $\vec q = \vec k_i - \vec k_f$. The Plane-Wave Impulse
Approximation (PWIA) amplitude is obtained from the 
expression~(\ref{eq:tdwiaprop})
by replacing the distorted proton waves by plane waves~:
\begin{eqnarray}
T^{DWIA}_{if} = 
\sum_{n \ell j m} \!<J_B j M_B m | J_A M_A>\! 
\sqrt {S_{n \ell j}} \times ~~~~~~~~~~~~~~~~~~~~~~~~~~ \nonumber \\
 \int d\vec R~
\exp( i \vec q \!\cdot\! \mu \vec R\,) 
\exp( -i \vec k_p \!\cdot\! \mu \vec R\,)~
\phi_{n \ell j m} (\vec R\,)~~~~~~~~~~~~~ \nonumber \\
=  \sum_{n \ell j m} \!<J_B j M_B m | J_A M_A>\! 
\sqrt {S_{n \ell j}}\int d\vec R~
\exp( -i \vec p_m \!\cdot\! \vec R\,)~
\phi_{n \ell j m} ( \vec R\,),
\label{eq:tdwiaprop2}
\end{eqnarray}
where in the last expression the proton momentum $\vec k_p$, which is 
the center-of-mass momentum, has been written in terms of laboratory 
momenta~: $\vec k_{p} = \vec k_p^{lab} + \vec p_m /(A-1)$.  
Expression~(\ref{eq:tdwiaprop2}) is just the Fourier transform of the 
BSWF. After including the angular momentum and spin parts and squaring 
the $T_{if}$-matrix element the well known expression~\cite{frul84} 
for the \eep\ cross section is obtained.

When distortions (of the proton or electron waves) are included, the 
cross section cannot be factorized any more into $\sigma_{ep}$ and 
S(\newem,\newpm) (see Eq.~(\ref{eq:sigdef})).  For convenience one 
defines the reduced cross section or distorted momentum distribution 
(both the experimental and the calculated one) by
\begin{equation}
\rho^{D}(\vec {p}_m) \delta(E_{m}-{E}_{tr}) =
S^{D}(E_{m},\vec {p}_m) =
\frac {1}{{k}\sigma_{ep}}
\frac {{\rm d}^6\sigma^{CDWIA}} {
{\rm d} E_{e\prime} {\rm d} \Omega_{e\prime} 
{\rm d} E_{p} {\rm d} \Omega_{p} }
\label{eq:rho}
\end{equation}
(compare eqs.  \ref{eq:sigdef} and \ref{eq:spfdef}).  In calculating 
the sixfold differential cross section there is some ambiguity in the 
current operator to be used as the proton is off-shell 
\cite{for83,naus90a,naus90b}.  However, in the used kinematics the 
influence on the spectroscopic factors of the different prescriptions 
as given by de Forest~\cite{for83} is smaller than a few percent.

%%%%%%%%%%%%%%%%%%%%%%%%%%%%%%%%%%%%%%%%%%%%%%%%%%%%%%%%%%%%%%%%%

\subsection{The \dhe\ reaction}

The basis of the DWBA description \cite{satch80,aust70,jack70} of 
transfer reactions $A+a \rightarrow B+b $, such as the \dhe\ reaction 
presently under study, is the transition amplitude~:
\begin{equation}
T_{\alpha\beta}\,=\,<\!\Psi^{(-)} | V_\alpha - U_\alpha |
\psi_A \psi_a \chi^{(+)}_\alpha(\vec k_\alpha,\vec r_\alpha)\!>,
\label{eq:transampl}
\end{equation}
where $\alpha~(\beta)$ is the entrance (exit) channel with projectile 
(ejectile) $a~(b)$ and target (final) nucleus $A~(B\,)$ and $\vec 
r_\alpha$ the displacement of $a$ from $A$ (see 
Fig.~\ref{fig:reaction}b).  The interaction $V_\alpha$ is the sum of 
two-body interactions between the nucleons of the projectile and those 
of the target nucleus.  The wave function $\psi_a~(\psi_A)$ is the 
internal wave function of the projectile (target nucleus), while 
$\chi^{(+)}_{\alpha}(\vec k_\alpha,\vec r_\alpha)$ is the solution of 
the Schr\"odinger equation for the incoming particle with the 
distorting potential $U_\alpha$, usually chosen to be an optical 
potential that fits the elastic scattering in channel $\alpha$, and 
$\Psi^{(-)}$ is the exact wave function of the system with 
incoming-wave boundary conditions.

In the DWBA method the following approximations are made~: \hfil\break 
First the exact wave function $\Psi^{(-)}$ is replaced by a product of 
internal wave functions of the outgoing particle $\psi_b$, the 
residual nucleus $\psi_B$ and a function $\chi^{(-)}_\beta(\vec 
k_\beta,\vec r_\beta)$ describing the elastic scattering of the 
outgoing particle off the final nucleus B. This leads to the 
distorted-wave transition amplitude~:
\begin{eqnarray}
T^{DW}_{\alpha\beta} & = &
<\!\chi^{(-)}_\beta(\vec k_\beta,\vec r_\beta)~\psi_b \psi_B
| V_\alpha - U_\alpha | \psi_A 
\psi_a~\chi^{(+)}_\alpha(\vec k_\alpha,\vec r_\alpha)\!> 
\nonumber \\
{~} & = &
\!\int\!d\vec r_\beta \int\!d\vec r_\alpha~
\chi^{(-)*}_\beta(\vec k_\beta,\vec r_\beta)
<\!B b | V_\alpha - U_\alpha | A a\!>
\chi^{(+)}_\alpha(\vec k_\alpha,\vec r_\alpha),
\label{eq:tdwba}
\end{eqnarray}
where $\vec r_\beta$ is the displacement of $b$ from $B$ (see 
Fig.~\ref{fig:reaction}a).  In the nuclear matrix element $<\!  \psi_b 
\psi_B | V_\alpha - U_\alpha | \psi_A \psi_a\!>$ the integration is 
performed over all coordinates independent of $\vec r_\alpha$ and 
$\vec r_\beta$.

The second approximation deals with the interaction $V_\alpha - 
U_\alpha$, which is replaced in the prior formalism by the interaction 
between the transferred nucleon and the projectile nucleons~:
\begin{equation}
V_\alpha - U_\alpha \approx \sum_i^a V_{in}\,=\,V_{an} (\vec r\,),
\label{eq:vint}
\end{equation}
where the sum runs over all the constituents of the projectile $a$, 
$n$ is the nucleon to be transferred and $\vec r\,$ is the 
displacement of the pickup nucleon from the center of mass of the 
projectile.  (In principle this interaction should be taken off-shell.  
In view of the many other approximations made, this point is probably 
not relevant and is always neglected in analyses.)

The above approximations are based on the following assumptions made 
for the transfer reaction mechanism itself.  Firstly it is assumed 
that the interaction that drives the reaction is weak enough so that 
the reaction process may be treated in first order perturbation 
theory.  Secondly the reaction is assumed to be a one-step process; 
the transferred nucleon is picked up by the incoming projectile, 
whereas all other target nucleons do not change their state of motion.

The distorted waves, $\chi^{(+)}_\alpha(\vec k_\alpha,\vec r_\alpha)$ 
and $\chi^{(-)}_\beta(\vec k_\beta,\vec r_\beta)$, are usually chosen 
to be the wave functions of optical potentials describing elastic 
scattering in the entrance and exit channels.  However, as already 
mentioned, the same elastic scattering data can be described with 
different parameterizations of the optical potential, resulting in 
waves identical at large distances but differing in the nuclear 
region.  Their contribution to the transition amplitude will differ 
accordingly and gives rise to an extra uncertainty in the 
spectroscopic factors deduced from transfer reactions.  In contrast to 
the \eep\ reaction where only one wave function enters that is 
generated in an optical model potential, in the \dhe\ reaction two 
such wave functions are entering.  Moreover, the uncertainty in the 
optical-model wave functions of composite particles in the interior of 
the nucleus is appreciably larger than that for nucleons.  
Consequently, the uncertainties due to different possible 
parameterizations of the optical-model potential are larger for 
spectroscopic factors deduced from the \dhe\ reaction.

As pointed out earlier, non-locality corrections must be applied to 
the wave functions obtained from the (local) optical potential.  This 
correction affects the wave functions of the projectile and ejectile 
in the region where the transfer takes place.

The nuclear matrix element in Eq.~(\ref{eq:tdwba}) can be expanded, 
along the lines given in \cite{jack70}, into the nuclear overlap 
integral as given in Eq.~(\ref{eq:overlapexp}) and into the overlap 
between the projectile and ejectile~: $ f(\vec r\,)~=~<\psi^*_b 
(\xi_a,\vec r\,) | V_{an}(\vec r\,) | \psi_a (\xi_a)>$.  This gives 
for the DWBA transition amplitude~:
\begin{equation}
T^{DW}_{\alpha\beta} \propto
\sum_{n \ell j m}\! \sqrt {S_{n \ell j}} \!
\int d\vec r_\beta \! \int d\vec r_\alpha~
\chi^{(-)*}_\beta(\vec k_\beta,\vec r_\beta)
\phi_{n \ell j m} (\vec R\,) f(\vec r\,)
\chi^{(+)}_\alpha(\vec k_\alpha,\vec r_\alpha).
\label{eq:tdwba2}
\end{equation}
The evaluation of this amplitude involves a six-dimensional integral 
over $\vec r_\alpha$ and $\vec r_\beta$.  A reduction to a more 
convenient three-dimensional integral is achieved in the zero-range 
approximation, where the effective interaction $V_{an} (\vec r\,)$ is 
assumed to have a range equal to zero, so that
\begin{eqnarray}
f(\vec r\,)\, & = & D_0 \delta (\vec r\,),  \nonumber \\
D_{0} & = & \int {\rm d} \vec r\ <\psi^*_b 
(\xi_a,\vec r\,) | V_{an}(\vec r\,) | \psi_a (\xi_a)>.
\label{eq:fzero}
\end{eqnarray}
The physical meaning of this approximation is that the ejectile $b$ is 
assumed to be emitted at the same position where the absorption of the 
projectile $a$ has taken place.  The effect of neglecting the finite 
range of the interaction is that the spectroscopic factors deduced 
from transfer reactions in a zero-range analysis are larger than the 
ones obtained from a full finite-range analysis \cite{kra88a}.

Full finite-range calculations are hard to perform because of the six 
dimensional integral in Eq.~(\ref{eq:tdwba2}).  However it has been 
shown by Buttle and Goldfarb \cite{buttle64} that the effects of the 
finite range of the interaction can be taken into account 
approximately (local energy approximation, LEA) by replacing the delta 
function in Eq.~\ref{eq:fzero} by the following radial factor~:
\begin{equation}
\Lambda (r) = [ 1 \!+\! {{2 m_a m_p} \over {\hbar^2 m_b}} R_{fr}^2
( {E}_\alpha - U_\alpha(\mu r)
+ {E}_p - U_p(r)
- {E}_\beta + U_\beta(r) )]^{-1},
\label{eq:ffinite}
\end{equation}
in which $R_{fr}$ is the finite range distance.

%%%%%%%%%%%%%%%%%%%%%%%%%%%%%%%%%%%%%%%%%%%%%%%%%%%%%%%%%%%%%%%%%

\section{BSWF probing functions}
\label{sec:bswf}

As shown in Section~\ref{sec:reactions}, the transition amplitudes of 
both the \eep\ and \dhe\ reaction consist of the nuclear matrix 
element sandwiched between the incoming and outgoing distorted 
"probing" waves~(see eqs.~(\ref{eq:tmat2}) and (\ref{eq:tdwba})).  In 
this section it is investigated to which part of the BSWF the \eep\ 
reduced cross section and the \dhe\ cross section are sensitive.

Cross sections are obtained from the T-matrix elements by integrating 
the radial coordinate from zero to infinity.  The radial sensitivity 
of the cross section was investigated by varying the lower radial 
integration bound between 0 and 10~fm and plotting these results as~:
\begin{equation}
P(r) = { 1 \over {\Delta r}} 
(\sigma_{r-\Delta r/2} - \sigma_{r+\Delta r/2}),
\label{eq:prob}
\end{equation}
where $\sigma_{x}$ reperesents the cross section calculated with a 
lower limit $x$ on the integral over the T-matrix element :
\begin{equation}
\sigma_{x} \propto | \int_{x}^\infty T(r)~{\rm d} r |^2.
\label{eq:radsens}
\end{equation}
In this way the separate contribution of the interval $\Delta r$ 
around $r$ to the cross section was obtained and hence the part of the 
BSWF to which the reaction is sensitive can be determined.

\begin{figure}
\centerline{\epsfysize=15cm \epsfbox{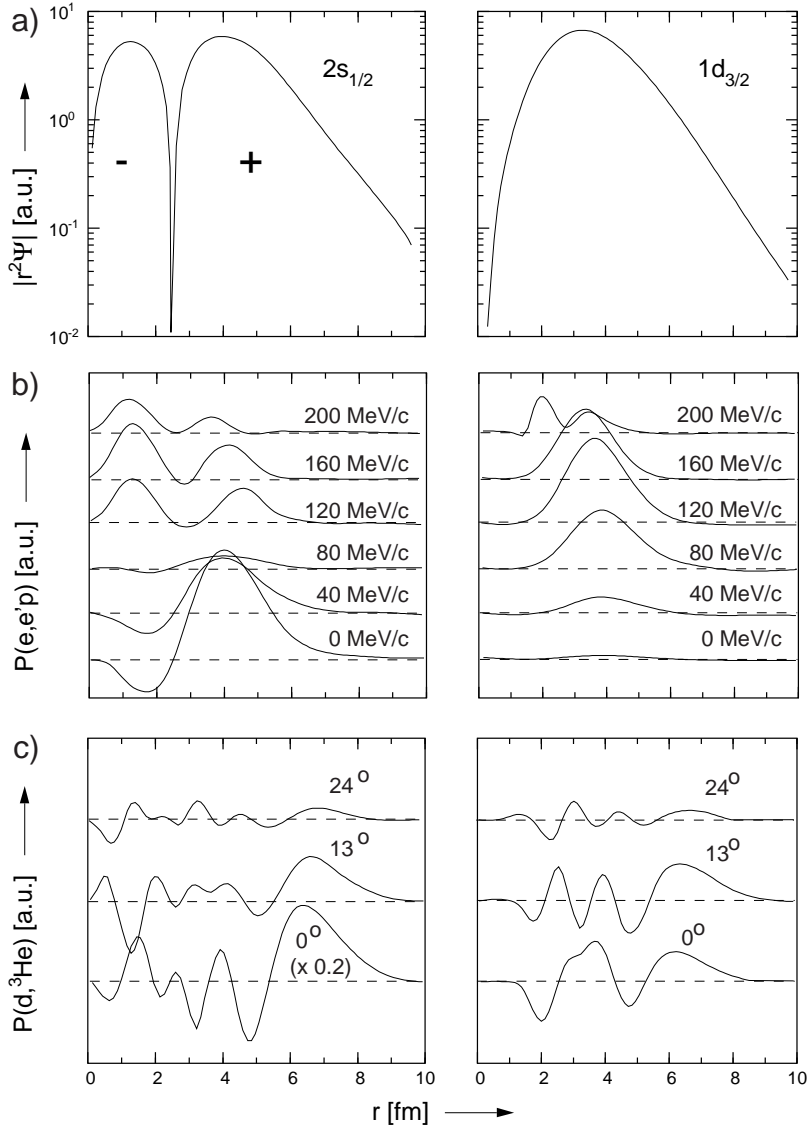} }
\caption{ a) The BSWF as obtained from the present \eep\ experiment 
for the \dthree\ and \twos\ orbitals.  b) The sensitivity P (see text) 
of the \eep\ momentum distributions to these BSWF. c) The sensitivity 
P (see text) of the \dhe\ differential cross sections to these BSWF. }
\label{fig:bswf}
\end{figure}

For the reaction \caeight\eep\ these calculations were performed for 
the transitions leading to the \halfpl\ ground-state and the first 
\threepl\ excited state in \kseven.  The BSWF shown in the upper part 
of Fig.~\ref{fig:bswf} that was used in these calculations was 
generated in a Woods-Saxon well with the parameters as given in 
Section~\ref{sec:cadata}.  In that section the other parameters that 
entered these CDWIA calculations are also given.  The results of the 
calculation of $P(r)$ for the \eep\ reaction are shown in the middle 
part of Fig.~\ref{fig:bswf} for different values of the missing 
momentum.  From this figure it can be seen that the \eep\ reaction is 
sensitive to the whole BSWF and the largest contribution to the 
momentum distribution comes from those regions in $r$, where also 
$r^2\Phi_{nlj}(r)$ is large.  For the \dthree\ orbital this is the 
region between r=2.5 and 4.5~fm and for the \twos\ orbital between 
r=0.7 and 6.7~fm.  Because of the node in the \twos\ orbital, for low 
missing momenta, there is a destructive contribution to the momentum 
distribution from the inner lobe, whereas this contribution becomes 
constructive for high missing momenta.

For the above mentioned transitions the sensitivity to the BSWF was 
also determined for the \dhe\ reaction.  The DWBA calculations were 
performed with the parameters as given in Section~\ref{sec:cadata} and 
the same BSWF as for the \eep\ experiment was used.  The results of 
these calculations are presented in the lower part of 
Fig.~\ref{fig:bswf}.  Here it can be seen that apart from strong 
interferences between the incoming and outgoing distorted waves in the 
interior of the nucleus, the \dhe\ reaction is most sensitive to the 
region between $r$=5 and 10~fm.  Therefore, the \dhe\ reaction is not 
sensitive to the details of the BSWF inside the nucleus.  In the 
region where the \dhe\ reaction is sensitive the BSWF has the global 
form~: $\nu\exp(-\kappa r)$, where $\kappa$ depends on the (measured) 
binding energy of the proton, and the normalization $\nu$ depends on 
the depth and shape of the potential that generates the BSWF. As the 
spectroscopic factor is the integral of the BSWF over the total radial 
region, one can only determine spectroscopic factors from the \dhe\ 
reaction by assuming some shape for the BSWF.

The conclusion is that with the \eep\ reaction the BSWF is probed in 
the whole radial region whereas, with the \dhe\ reaction only the 
exponential tail of the BSWF is probed.  This tail is very sensitive 
to the exact shape of the used proton-binding potential.  The shape of 
the BSWF introduces thus a large model dependence, sometimes up to 
50\% \cite{kra88a}, in spectroscopic factors deduced from \dhe\
experiments.

Given this sensitivity, it is even questionable whether ratios of 
spectroscopic factors for different isotopes can be determined 
accurately in the \dhe\ reaction~\cite{wag85}, as it is not certain 
that the radius of the BSWF well scales with A$^{1/3}$.

%%%%%%%%%%%%%%%%%%%%%%%%%%%%%%%%%%%%%%%%%%%%%%%%%%%%%%%%%%%%%%%%%

\section{Analysis of the \caeight\ data}
\label{sec:cadata}

\subsection{Analysis of the \caeight\eep\ experiment}

\begin{figure}
\centerline{\epsfysize=10cm \epsfbox{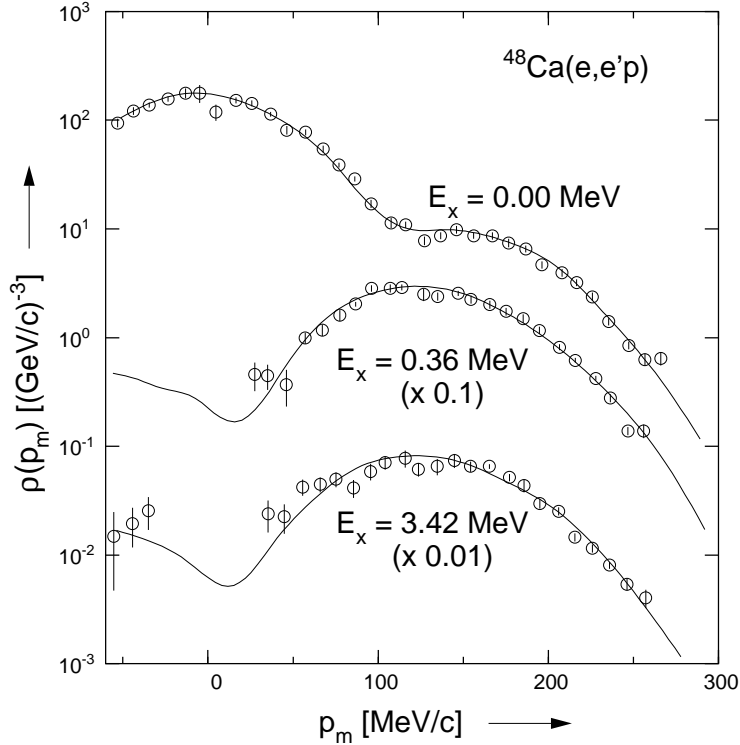} }
\caption{ Momentum distributions for the first three positive-parity 
transitions in the reaction \caeight\eep\ together with curves 
obtained from the \mbox{CDWIA} calculations as mentioned in the text.  
}
\label{fig:momdis}	
\end{figure}

The experimental \eep\ momentum distributions were obtained with the 
coincidence set-up at NIKHEF \cite{vries84}.  Two metal foils with a 
thicknesses of 7.3 and 15.0~mg/cm$^2$, enriched to 95.2~\% in
\caeight\ were used.  Reduced cross sections were obtained under 
parallel kinematic conditions in the range between -60 and 260~MeV/c.  
The electron beam energy was 440~MeV and the outgoing proton kinetic 
energy was 100~MeV. The experimental systematic error on the extracted 
distributions is 4~\%.  Further details can be found in Ref.
\cite{kra88b}.  The CDWIA calculations were performed with the code 
DWEEPY~\cite{gius88}.  The proton optical-potential parameters were 
obtained from the work of Schwandt et al.  \cite{schw82}.  A 
non-locality correction according to the prescription of Perey 
\cite{perey63} (see also Eq.~(\ref{eq:perey})) was applied with a 
range parameter $\beta$ of 0.85~fm.  The bound state wave function was 
calculated in a Woods-Saxon well with a diffuseness $a_0$ of 0.65~fm 
and a Thomas spin-orbit parameter $\lambda$ of 25.  A non-locality 
correction was also applied to the BSWF with a $\beta_{nloc}$ of 
0.85~fm.  The well depth $V_0$ and the radius parameter r$_0$ were 
adjusted with the separation energy as a constraint to get the best 
description of the measured momentum distributions.  In 
Fig.~\ref{fig:momdis} these reduced cross sections are shown for the 
transitions to the first three positive parity states together with 
the results of the CDWIA calculations, while in 
Table~\ref{tab:caspecfac} the deduced spectroscopic factors and radii 
of the BSWF are given.

\begin{table}
\caption[\caeight\ spectroscopic factors.]
{ Spectroscopic factors for proton pick up from \caeight\ deduced from
	\eep\ and from \dhe\ experiments.}
  \begin{tabular}{|c|c|c|c|c|c|c|c|} \hline
	$E_x$ & $J^\pi$ & $r_0$ & $r_{\rm rms}$~$^{+)}$ & $S$ \eep & 
	\multicolumn{2}{c|}{S\dhe\cite{banks85}} & $S$\dhe\cite{mats84} \\
	{[MeV]} & & [fm] & [fm] & & LZR & NLFR & NLFR \\ \hline
	0.00 & $1/2^+$ & 1.228(47) & 3.58(10) & 1.07~(~7) & 1.55 & 0.96 & 
	0.94(25) \\
	0.36 & $3/2^+$ & 1.254(48) & 3.54(10) & 2.26~(16) & 4.16 & 2.39 & 
	2.31(65) \\
	3.42 & $5/2^+$ & 1.128(44) & 3.39(~9) & 0.683(49) & 1.02 & 1.28 & 
	1.07(31) \\
	3.85 & $1/2^+$ & 1.294(51) & 3.59(10) & 0.167(14) & 0.28 & 0.12 & 
	\\
	3.95 & $3/2^+$ & 1.288~~~~~ & 3.54~~~~~ & 0.323(27) & 0.70 & 0.32 
	& \\
	5.24 & $5/2^+$ & 1.192(48) & 3.49(~8) & 0.288(21) & 0.32 & 0.27 & 
	\\
	5.49 & $5/2^+$ & 1.182(46) & 3.47(~9) & 0.746(52) & 0.94 & 0.84 & 
	\\
	6.51 & $5/2^+$ & 1.265(56) & 3.62(12) & 0.160(14) & 0.22 & 0.11 & 
	\\
	6.87 & $5/2^+$ & 1.162(65) & 3.41(14) & 0.070(~7) & 0.14 & 0.14 & 
	\\
	7.81 & $5/2^+$ & 1.243(49) & 3.56(~9) & 0.434(32) & 0.71 & 0.42 & 
	\\
	8.13 & $5/2^+$ & 1.299(54) & 3.46(12) & 0.228(19) & 0.33 & 0.26 & 
	\\
\hline
	\multicolumn{8}{l}{\footnotesize $^{+)}$ rms radius in the proton 
	A-1 system.}
  \end{tabular}
\label{tab:caspecfac}
\end{table}
%%%%%%%%%%%%%%%%%%%%%%%%%%%%%%%%%%%%%%%%%%%%%%%%%%%%%%%%%%%%%%%%%

\subsection{Re-analysis of the \caeight\dhe\ experiments}

\begin{figure}
\centerline{\epsfysize=7.5cm \epsfbox{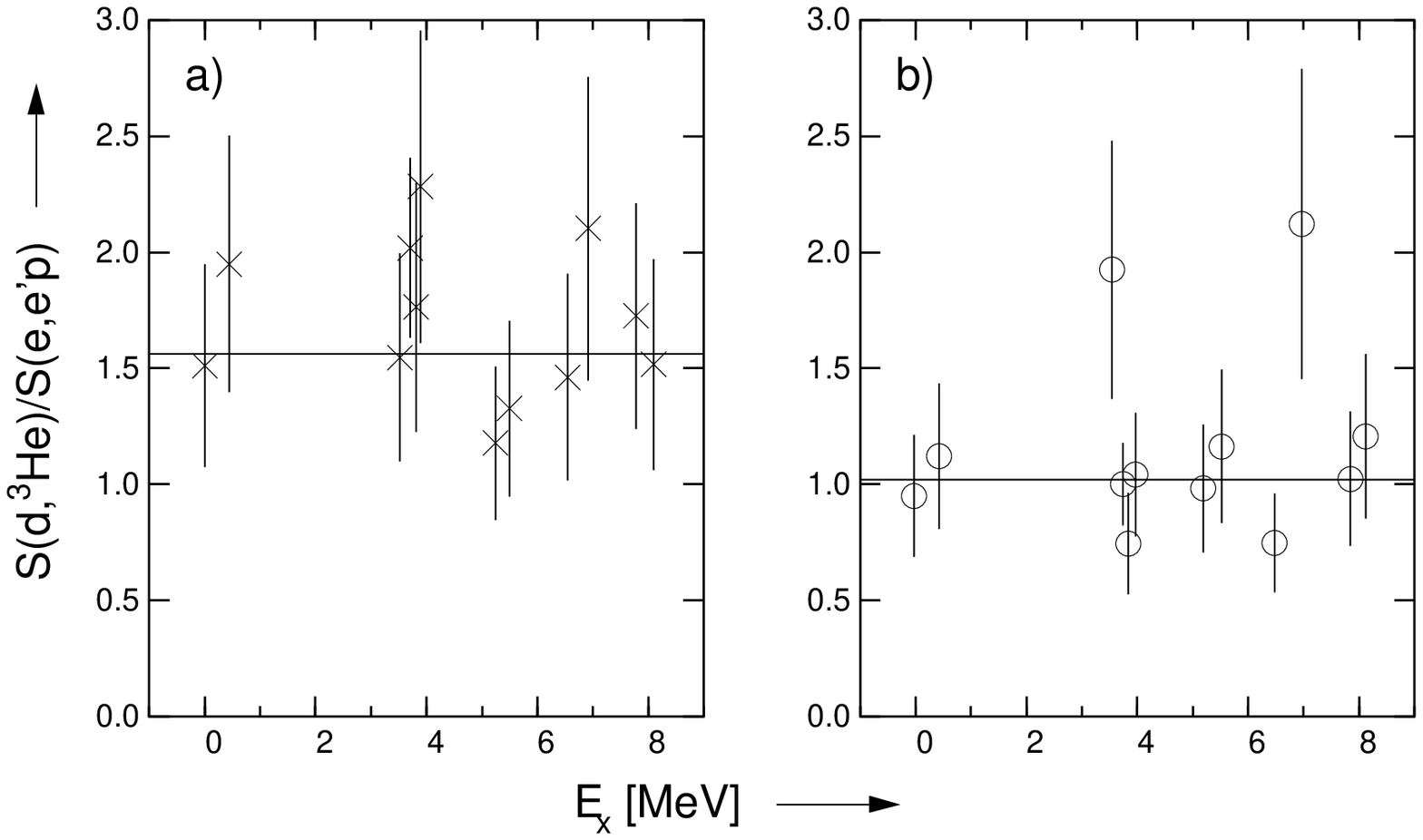} }
\caption{ a) The ratio of the spectroscopic factors given in 
Ref.~\cite{banks85} derived from a local zero-range analysis of the 
\caeight\dhe\ experiment and those obtained from the present 
\caeight\eep\ experiment for various transitions observed in both 
reactions as a function of excitation energy.  The solid line 
represents the average.  b) The same ratio after including 
non-locality and finite-range corrections in the \dhe\ analysis and 
using the BSWF's obtained from the \eep\ experiment.  A 30\% error was
assigned to the \dhe\ spectroscopic factors in both cases.  }
\label{fig:ratio}	
\end{figure}

For the comparison with the data from the \eep\ experiment a 
re-analysis of two ($\vec {\rm d},^3$He) experiments was performed.  
The first \dhe\ experiment~\cite{banks85} was performed with an 
incoming deuteron energy of 79.2~MeV. Further details of this 
experiment can be found in the original paper~\cite{banks85}.  In the 
analysis in that paper a local zero-range DWBA calculation was used 
for the extraction of the spectroscopic factors together with a BSWF 
potential well with $r_0$=1.25 fm, $a_0$=0.60 fm and $V_0$ adjusted to 
get the correct binding energy.  Non-locality corrections were not 
applied to the BSWF. The ratio of the spectroscopic factors given 
in~\cite{banks85} to those deduced from the present \eep\ experiment 
for several discrete states is shown in Fig.~\ref{fig:ratio}a as a 
function of the excitation energy.  The \dhe\ spectroscopic factors 
calculated this way are on the average 50\% higher than those obtained
from the \eep\ experiment.

In the present re-analysis, performed with the code DWUCK4 
\cite{kunzpri}, non-locality corrections and finite range corrections 
via the LEA approach were included together with the BSWF obtained 
from the present \eep\ experiment.  The same optical model parameter 
sets for the deuteron and $^3$He waves were used as in Ref.  
\cite{banks85}.  Spectroscopic factors deduced with this re-analysis 
are given in Table~\ref{tab:caspecfac}.  In order to estimate 
realistic errors on the spectroscopic factors deduced from the \dhe\ 
experiment the following sources of uncertainties were taken into 
account~: 
i) a total experimental systematic error of 5\% which includes the
error on the target thickness;
ii) the effect of the uncertainty (about 3\%) in the rms radii 
obtained from
   the \eep\ experiment, which yields
   25\% for the transition to the \halfpl,
   28\% for the transition to the \threepl\ and
   29\% for the transition to the \fivepl state;
iii) a 10\% uncertainty due to the $<{\rm d}|^3$He$>$ overlap 
function (the value of $D_0$); 
and 
iv) at least a 10\% uncertainty due to different possible 
parameterizations of the deuteron and $^3$He optical potentials.  The 
latter number is taken from~\cite{kra88a}, where the sensitivity of 
the spectroscopic factors to different optical potential 
parameterizations was investigated for the reaction $^{51}$V\dhe 
$^{50}$Ti at 53 MeV.

The ratio of the spectroscopic factors obtained from the present 
analysis of the \dhe\ data to those obtained from the \eep\ experiment 
is plotted in Fig.~\ref{fig:ratio}b.  The average ratio is one, so it 
is concluded that, except for two points, there is a very good 
agreement between the spectroscopic factors obtained from both 
reactions.  This agreement is obtained by including non-locality and 
finite-range corrections in the analysis together with experimental 
BSWF's obtained from the \eep\ experiment.  Finite-range corrections 
reduce the spectroscopic factors by about 15~\% and the use of the 
BSWF determined from the \eep\ analysis gives a a further reduction of 
30 to 40~\%.

The deviation of the spectroscopic factors and the relatively small 
rms radius for the transition leading to the 3.42~MeV excited state in 
\kseven\ might be ascribed to some unresolved \newfseven\ strength at 
3.4~MeV \cite{kra88b} but in the scarce literature on the level scheme 
of \kseven\ no \sevenmin\ states have been reported so far.  The 
reduced cross section for this transition can also be described well 
with a BSWF with an rms radius of 3.47~fm, which is the average value 
for the \newdfive\ orbital obtained from all the \fivepl\ transitions 
observed in the \eep\ experiment.  This gives a 3~\% lower
spectroscopic factor in the \eep\ experiment, but in the \dhe\ 
analysis the spectroscopic factor drops by 14~\%.  The deviation for
the spectroscopic factor for the very weak transition at 6.87~MeV may 
be due to the uncertainty in the rms radius that is not well 
determined from the \eep\ experiment.  It is also possible that 
two-step processes have a different effect on the \eep\ and \dhe\ 
cross sections for this weak transition.

\begin{figure}
\centerline{\epsfysize=10cm \epsfbox{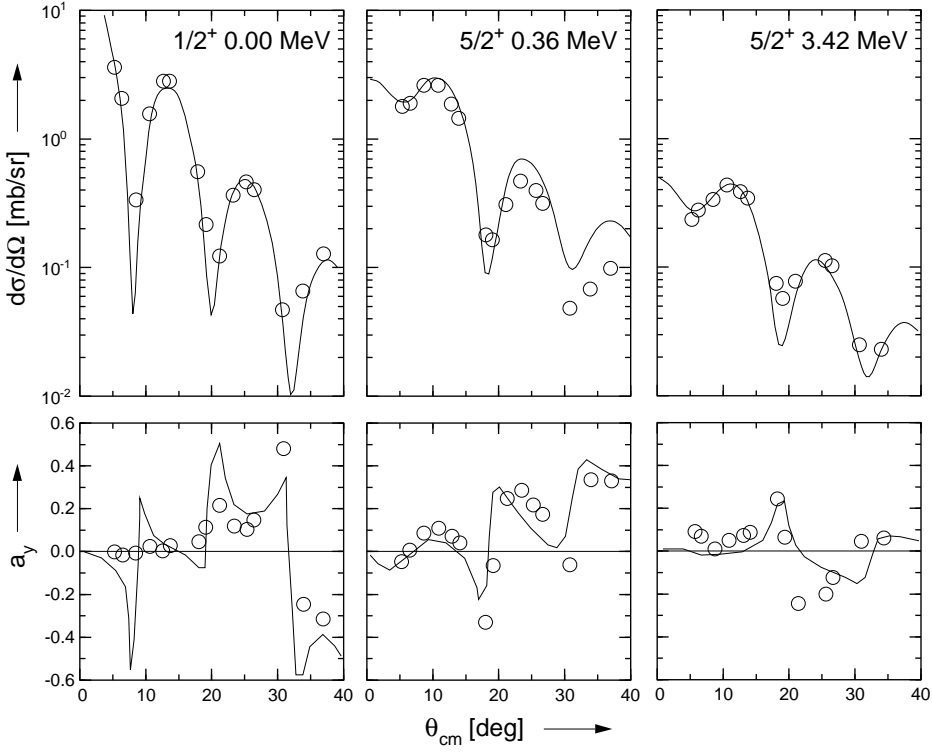} }
\caption{ Angular distributions and asymmetries for the transitions 
leading to the first three positive-parity states in \kseven\ in the 
reaction \caeight\dhe\ at $E_d$ = 56 MeV. }
\label{fig:angdis}
\end{figure}

The second experiment was performed with an incoming deuteron energy 
of 56~MeV \cite{mats84}.  Angular distributions and asymmetries were 
measured for the first three positive parity transitions, leading to 
\halfpl, \threepl and \fivepl\ states in \kseven, see 
Fig.~\ref{fig:angdis}.  The used optical-model potential 
parameterizations for the deuteron and $^3$He waves were obtained from 
elastic deuteron and $^3$He scattering off \caeight\ \cite{matsp} and 
are listed in Table~\ref{tab:optpot}.  The non-locality corrections 
were taken into account according to the prescription of Perey 
\cite{perey63} (see also Eq.~(\ref{eq:perey})).  For the deuteron and 
$^3$He wave-functions the parameters are given in the last column of 
Table~\ref{tab:optpot}.  The same BSWF as obtained in the analysis of 
the \eep\ experiment was employed in the calculation of the \dhe\ 
cross sections.

\begin{table}
\caption[Optical potentials for \caeight]
{Optical potential parameters \cite{matsp} for deuteron and $^3${\rm He}
	scattering off $\caeight$.}

  \begin{tabular}{|c|c|c|c|c|c|c|c|c|} \hline
	particle & $E_{\rm lab}$ & $V_r$ & $r_0$ & $a_0$ & $W_v$ & $W_d$ & 
	$r_i$ & $a_i$ \\
		 & [MeV] & [MeV] & [fm] & [fm] & [MeV] & [MeV] & [fm] & [fm] 
		 \\
\hline
	d & 56.0 & 79.0 & 1.154 & 0.755 & 6.431 & 4.992 & 1.490 & 0.694 \\
	$^3$He & 45.0 & 184.6 & 1.115 & 0.713 & - & 21.95 & 1.217 & 0.812 
	\\
\hline
 \end{tabular}
  \begin{tabular}{|c|c|c|c|c|c|} \hline
particle & $V_{ls}$ & $r_{ls}$ & $a_{ls}$ & $r_c$ & $\beta_{\rm nloc}$ 
\\
	 & [MeV] & [fm] & [fm] & [fm] & [fm] \\ \hline
d & 7.60 & 0.986 & 0.777 & 1.30 & 0.54 \\
$^3$He & - & - & - & 1.40 & 0.2~~ \\ \hline
 \end{tabular}
\label{tab:optpot}
\end{table}

Finite-range effects were included by applying the LEA correction 
(Eq.~(\ref{eq:ffinite})).  A finite range distance of 0.77~fm was used 
together with the Bassel normalization \cite{bassel66} $D_0$=2.95 for 
the overlap between the deuteron and the $^3$He ejectile.  The LEA 
approach was compared to full finite-range calculations performed with 
the code DWUCK5 \cite{kunzpri}.  In the latter calculations the 
D-state of the deuteron was included and a \dhe\ overlap function was 
used that yields $D_0$=2.95.  The cross sections in the full finite 
range calculations were globally 10 to 15~\% larger (and hence the
spectroscopic factors smaller) than in the LEA calculation.  As the 
used value of $D_0$ has also an uncertainty for convenience all DWBA 
calculations to be presented were performed in LEA and the presented 
spectroscopic factors were obtained from those.  In 
Fig.~\ref{fig:angdis} the results of DWBA calculations are shown for 
the transitions mentioned.  Both angular distributions and asymmetries 
are described well with the used optical potentials and the BSWF's 
obtained from the \eep\ experiment.  The spectroscopic factors 
extracted for these three transitions are given in 
Table~\ref{tab:caspecfac}.  An estimate of the errors on the 
spectroscopic factors from this \dhe\ experiment gives typically 
numbers between 25 and 35~\%.

%%%%%%%%%%%%%%%%%%%%%%%%%%%%%%%%%%%%%%%%%%%%%%%%%%%%%%%%%%%%%%%%%
\begin{figure}
\centerline{\epsfysize=7.5cm \epsfbox{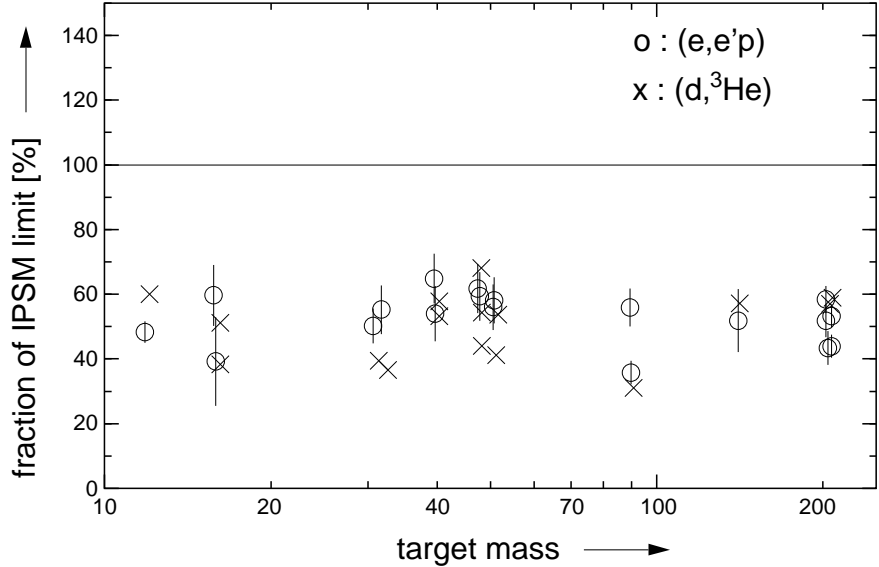} }
\caption{ Summed spectroscopic strength of the valence orbit as 
function of the mass number normalized to the Independent-Particle 
Shell Model limit according to literature values from \eep\ 
experiments and from a re-analysis, as discussed in the text, of the 
\dhe\ data presented in Fig.~\ref{fig:saval}.  }
\label{fig:savalboth}
\end{figure}

\begin{table}
  \caption[spectroscopic factors.]
  {Spectroscopic factors deduced from the re-analysis of existing 
  \dhe\ data.} \begin{tabular}{|c|c|r|l|c|l|l|} \hline
target & $E_x$ & $J^\pi$~~ & ~~$S$ \eep\ & $r_0$ & ~$S$ \dhe\ & ~$S$ 
\dhe\ \\
nucleus & [MeV] & {~} & {~} & [fm] & ~literature~ & ~re-analysis~ \\
\hline
$^{12}$C & 0.000 & 3/2$^-$ & 1.72~(11)~\cite{ste87} & 1.35~(2) & 
2.98~\cite{mair75} & 1.72~{~} \\
{~} & 2.125 & 1/2$^-$ & 0.26~(~~2)~{~} & 1.65~(2) & 0.69 & 0.27~{~} \\
{~} & 5.020 & 3/2$^-$ & 0.20~(~~2)~{~} & 1.51~(2) & 0.31 & 0.11~{~} \\
$^{16}$O & 0.000 & 1/2$^-$ & 1.27~(13)~\cite{leuspri} & 1.37~(3) & 
2.30~\cite{hieb67} & 1.02~{~} \\
{~} & 6.320 & 3/2$^-$ & 2.25~(22)~{~} & 1.28~(2) & 3.64 & 1.94~{~} \\
$^{31}$P & 0.000 & ~~0$^+$ & 0.40~(~~3)~\cite{wespri} & 1.27~(2) & 
0.62~\cite{mack74} & 0.36~{~} \\
{~} & 2.239 & ~~2$^+$ & 0.60~(~~5)~{~} & 1.18~(3) & 0.72 & 0.49~{~} \\
{~} & 3.498 & ~~2$^+$ & 0.28~(~~2)~{~} & 1.12~(3) & 0.30 & 0.19~{~} \\
$^{40}$Ca & 0.000 & 3/2$^+$ & 2.58~(19)~\cite{kram90} & 1.30~(5) & 
3.70~\cite{doll76} & 2.30~{~} \\
{~} & 2.522 & 1/2$^+$ & 1.03~(~~7)~{~} & 1.28~(6) & 1.65 & 1.03~{~} \\
$^{51}$V & 0.000 & 7/2$^-$ & 0.37~(~~3)~\cite{her88} & 1.30~(3) & 
0.73~\cite{hint67} & 0.30~\cite{kra88a} \\
{~} & 1.554 & 7/2$^-$ & 0.16~(~~2)~{~} & 1.31~(4) & 0.39 & 0.15 \\
{~} & 2.675 & 7/2$^-$ & 0.33~(~~3)~{~} & 1.32~(3) & 0.64 & 0.26 \\
{~} & 3.199 & 7/2$^-$ & 0.49~(~~4)~{~} & 1.34~(3) & 1.05 & 0.39 \\
{~} & 4.410 & 1/2$^+$ & 0.28~(~~3)~{~} & 1.22~(3) & 0.63 & 0.22 \\
{~} & 6.045 & 1/2$^+$ & 0.35~(~~3)~{~} & 1.27~(4) & 1.10 & 0.30 \\
$^{90}$Zr & 0.000 & 1/2$^-$ & 0.72~(~~7)~\cite{her88} & 1.32~(3) & 
1.80~\cite{stuir80} & 0.60~\cite{wagpri} \\
{~} & 0.909 & 9/2$^+$ & 0.54~(~~5)~{~} & 1.31~(2) & 1.25 & 0.30 \\
{~} & 1.507 & 3/2$^-$ & 1.86~(14)~{~} & 1.27~(2) & 3.90 & 1.20 \\
{~} & 1.745 & 5/2$^-$ & 2.77~(19)~{~} & 1.30~(2) & 8.90 & 2.40 \\
$^{142}$Nd & 0.000 & 5/2$^+$ & 1.39~(23)~\cite{lanen90} & 1.29~(9) & 
2.53~\cite{baer69} & 1.25~{~} \\
{~} & 0.145 & 7/2$^+$ & 3.14~(43)~{~} & 1.26~(8) & 6.28 & 3.79~{~} \\
{~} & 1.118 & 11/2$^-$ & 0.56~(~~7)~{~} & 1.28~(8) & 0.74 & 0.36~{~} 
\\
{~} & 1.300 & 1/2$^+$ & 0.05~(~~1)~{~} & 1.26~(9) & 0.11 & 0.07~{~} \\
\hline
  \end{tabular}
\label{tab:allspecfac}
\end{table}

\begin{table}
  \caption[spectroscopic factors.]
  {Spectroscopic factors (continued from Table~\ref{tab:allspecfac})} 
  \begin{tabular}{|c|c|r|l|c|l|l|} \hline
target & $E_x$ & $J^\pi$~~ & ~~$S$ \eep\ & $r_0$ & ~$S$ \dhe\ & ~$S$ 
\dhe\ \\
nucleus & [MeV] & {~} & {~} & [fm] & ~literature~ & ~re-analysis~ \\
\hline
$^{206}$Pb & 0.000 & 1/2$^+$ & 0.68~(~~6)~\cite{quint88} & 1.23~(9) & 
1.15~\cite{radha88} & 1.03~{~} \\
{~} & 0.203 & 3/2$^+$ & 1.10~(~~9)~{~} & 1.27~(9) & 1.77 & 0.99~{~} \\
{~} & 0.616 & 5/2$^+$ & 0.32~(~~3)~{~} & 1.23~(8) & 0.52 & 0.44~{~} \\
{~} & 1.151 & 3/2$^+$ & 0.52~(~~5)~{~} & 1.28~(9) & 0.66 & 0.37~{~} \\
{~} & 1.479 & 11/2$^-$ & 3.58~(32)~{~} & 1.25~(9) & 6.94 & 5.21~{~} \\
$^{208}$Pb & 0.000 & 1/2$^+$ & 0.98~(~~9)~\cite{quint88} & 1.25~(8) & 
1.8~~~\cite{lang84} & 1.5~~{~} \\
{~} & 0.350 & 3/2$^+$ & 2.31~(22)~{~} & 1.23~(8) & 3.8~ & 2.2~~{~} \\
{~} & 1.350 & 11/2$^-$ & 6.85~(68)~{~} & 1.16~(9) & 7.7~ & 5.4~~{~} \\
{~} & 1.670 & 5/2$^+$ & 2.93~(28)~{~} & 1.19~(8) & 3.5~ & 3.1~~{~} \\
{~} & 3.470 & 7/2$^+$ & 2.06~(20)~{~} & 1.15~(9) & 3.5~ & 2.9~~{~} \\
\hline
  \end{tabular}
\end{table}

\section{\dhe\ spectroscopic factors for other nuclei}
\label{sec:reanalysis}

A similar comparison as for \caeight\ has also been made for other 
nuclei where good \eep\ and \dhe\ data exist.  For these nuclei the 
\dhe\ data were re-analyzed in the same way as described above.  The 
optical potentials were taken from the original papers, non-locality 
and finite-range corrections were included in the same way as for 
\caeight\ and the BSWF's were taken from the \eep\ work.  Only pick-up 
from the valence shells was considered.  The result of this comparison 
is presented in Table~\ref{tab:allspecfac}, while in 
Fig.~\ref{fig:savalboth} the spectroscopic factors expressed as a 
fraction of the IPSM limit are shown.  The agreement between the 
spectroscopic factors for these transitions from the \dhe\ experiments 
and \eep\ experiments is very good.  The average ratio of \dhe\ over 
\eep\ spectroscopic factors is 1.01 with a spread of 0.25.  The error 
on the spectroscopic factors obtained from the \eep\ experiments, 
typically 10\%, was taken from their respective
references and a 25\% error was assigned to the spectroscopic factors
from the \dhe\ experiments.  The latter error is mainly due to to the 
large dependence of the spectroscopic factors on the rms radius of the 
BSWF~: $ {\Delta S / S } \approx 7 {\Delta r_{rms} / r_{rms} }$.

In view of the above conclusion that the spectroscopic factors deduced 
from \eep\ and \dhe\ reactions are in agreement, and exhaust only about 
60 \% of the IPSM value, the question may be asked why previously
applied \cite{CleP77} spin-dependent sum rules for hadronic transfer 
reactions yielded values close to 100 \%.  As argued earlier by us
\cite{Her88b}, this sum rule, which connects stripping and pick-up 
strengths, is only valid if all strength for a given spin is included 
in the summation.  This condition is clearly not fulfilled, as is 
known both from experimental data and from modern nuclear-structure 
calculations.  In angular-momentum decompositions 
\cite{her88,leuspri,wespri,kram90,quint88} of spectral functions 
obtained with the reaction \eep, it has been shown that spectroscopic 
strength distributions for a given angular momentum possess long tails 
extending to large energies.  These tails, which may contain up to 20 
\% of the total strength in the energy distribution, have not been
included in the sum-rule analysis.  Moreover, calculations of 
correlated nuclear matter \cite{pan84} have shown that not only the 
energy distributions of hole states, but also those of particle states 
exhibit such tails, which extend to several hundred MeV beyond the 
quasi-particle pole.  Consequently, application of the sum rule will 
necessarily fail, since appreciable parts of both the hole and the 
particle strength are lacking in the summation.

%%%%%%%%%%%%%%%%%%%%%%%%%%%%%%%%%%%%%%%%%%%%%%%%%%%%%%%%%%%%%%%%%

\section{Conclusion}
\label{sec:conclusion}

In this article it has been shown that spectroscopic factors obtained 
from \eep\ and \dhe\ experiments are mutually consistent, provided 
that in the DWBA calculations for the analysis of the \dhe\ data 
non-locality and finite-range corrections are included together with 
the BSWF obtained from \eep\ experiments.  It was also shown that the 
\eep\ reaction is sensitive to the whole BSWF, whereas the \dhe\ 
reaction is only sensitive to the exponential tail of the BSWF. This 
tail is very sensitive to the assumed shape of the potential well used 
to generate the BSWF. From the consistency of the obtained results it 
can be concluded that the reaction mechanism for these transitions in 
\eep\ as well as in the \dhe\ reaction is understood well enough to 
obtain reliable nuclear structure information.  From both reactions a 
spectroscopic strength of about 50 to 70\% of the IPSM limit is found
for strong valence transitions.

\begin{ack}
We would like to thank Dr.~N.~Matsuoka for making available to us the 
the \caeight\dhe\ data and the deuteron and $^3$He optical potentials, 
and Dr.  P.D. Kunz for many useful discussions on finite-range 
corrections and making available the codes DWUCK4 and DWUCK5.  This 
work is part of the research program of the National Institute for 
Nuclear Physics and High-Energy Physics (NIKHEF), which is made 
possible by the financial support from the Foundation for Fundamental 
Research of Matter (FOM) and the Netherlands Organization for the 
Advancement of Research (NWO).
\end{ack}

%%%%%%%%%%%%%%%%%%%%%%%%%%%%%%%%%%%%%%%%%%%%%%%%%%%%%%%%%%%%%%%%%

\end{document}